\documentclass[twocolumn]{aastex631}

\usepackage{hyperref}
\usepackage{amsmath} 
\usepackage{CJKutf8}
\usepackage{threeparttable}
\usepackage{makecell}
\usepackage{booktabs}

\usepackage{soul}

\hypersetup{hypertex=true,colorlinks=true,
linkcolor=blue,anchorcolor=blue,citecolor=blue}

\begin{document}

\title{Measurement of Neutral Atmosphere Density During the Years of Increasing Solar Activity Using \textit{Insight}-HXMT Data with the Earth Occultation Technique}

\correspondingauthor{Xiao-Bo LI, Yun-Wei YU}
\email{lixb@ihep.ac.cn, yuyw@ccnu.edu.cn}

\author[0009-0000-8848-1803]{Hao-Hui Zhang\begin{CJK*}{UTF8}{gbsn} (张浩辉) \end{CJK*}}
\affiliation{Institute of Astrophysics, Central China Normal University, Wuhan 430079, China}
\affiliation{Key Laboratory for Particle Astrophysics, Institute of High Energy Physics, Chinese Academy of Sciences, 19B Yuquan Road, Beijing 100049, China}
\author{Wang-Chen Xue\begin{CJK*}{UTF8}{gbsn} (薛王陈) \end{CJK*}} 
\affiliation{Key Laboratory for Particle Astrophysics, Institute of High Energy Physics, Chinese Academy of Sciences, 19B Yuquan Road, Beijing 100049, China}
\affiliation{University of Chinese Academy of Sciences, Chinese Academy of Sciences, Beijing 100049, China}

\author{Xiao-Bo Li\textsuperscript{*}\begin{CJK*}{UTF8}{gbsn} (李小波) \end{CJK*}} 
\affiliation{Key Laboratory for Particle Astrophysics, Institute of High Energy Physics, Chinese Academy of Sciences, 19B Yuquan Road, Beijing 100049, China}
 \author{Shuang-Nan Zhang\begin{CJK*}{UTF8}{gbsn} (张双南) \end{CJK*}} 
\affiliation{Key Laboratory for Particle Astrophysics, Institute of High Energy Physics, Chinese Academy of Sciences, 19B Yuquan Road, Beijing 100049, China}
  \author{Shao-Lin Xiong\begin{CJK*}{UTF8}{gbsn} (熊少林) \end{CJK*}} 
\affiliation{Key Laboratory for Particle Astrophysics, Institute of High Energy Physics, Chinese Academy of Sciences, 19B Yuquan Road, Beijing 100049, China}
   \author{Yong Chen\begin{CJK*}{UTF8}{gbsn} (陈勇) \end{CJK*}} 
\affiliation{Key Laboratory for Particle Astrophysics, Institute of High Energy Physics, Chinese Academy of Sciences, 19B Yuquan Road, Beijing 100049, China}
    \author{Hai-Tao Li\begin{CJK*}{UTF8}{gbsn} (李海涛) \end{CJK*}} 
\affiliation{University of Chinese Academy of Sciences, Chinese Academy of Sciences, Beijing 100049, China}
\affiliation{National Space Science Center, Chinese Academy of Sciences, Beijing, 100190, China }
    \author{Li-Ming Song\begin{CJK*}{UTF8}{gbsn} (宋黎明) \end{CJK*}} 
\affiliation{Key Laboratory for Particle Astrophysics, Institute of High Energy Physics, Chinese Academy of Sciences, 19B Yuquan Road, Beijing 100049, China}
     \author{Ming-Yu Ge\begin{CJK*}{UTF8}{gbsn} (葛明玉) \end{CJK*}} 
\affiliation{Key Laboratory for Particle Astrophysics, Institute of High Energy Physics, Chinese Academy of Sciences, 19B Yuquan Road, Beijing 100049, China}
      \author{Hai-Sheng Zhao\begin{CJK*}{UTF8}{gbsn} (赵海升) \end{CJK*}} 
\affiliation{Key Laboratory for Particle Astrophysics, Institute of High Energy Physics, Chinese Academy of Sciences, 19B Yuquan Road, Beijing 100049, China}
\author[0000-0002-1067-1911]{Yun-Wei Yu\textsuperscript{*}\begin{CJK*}{UTF8}{gbsn} (俞云伟) \end{CJK*}} 
\affiliation{Institute of Astrophysics, Central China Normal University, Wuhan 430079, China}


\begin{abstract}
The density of the Earth's middle and upper atmosphere is an important question in Earth science and is a critical factor in the design, operation, and orbital determination of low Earth orbit spacecraft. In this study, we employ the Earth Occultation Technique (EOT) combined with Maximum Likelihood Estimation to estimate the neutral atmospheric density by modeling the attenuation of X-ray photons during the occultation process of \textit{Insight}-HXMT observations of Crab Nebula. 
Based on 83 occultation datasets of the Crab Nebula observed by all three sets of telescopes of \textit{Insight}-HXMT between 2022 and 2024, we derived the atmospheric densities at altitudes ranging from 55\,--130\,km. We find a general agreement between our results and the prediction by the NRLMSIS model within the altitude ranges of  65\,-- 90\,km, 95\,--100\,km and 120\,--130\,km, 
particularly during periods of enhanced solar activity. 
However, we also find that the NRLMSIS model overestimates atmospheric density at altitudes 90\,--95\,km and 100\,--120\,km by approximately 20\%. Furthermore, since the atmospheric density measurements at altitudes of 55\,--\,65\,km may be subject to selection bias, we do not report the prediction accuracy of the NRLMSIS model at this altitude.

\end{abstract}


\section{Introduction}

\label{introduciton}
The Earth Occultation Technique (EOT) has been widely employed for satellite imaging and measuring the energy spectra and fluxes of high-energy astrophysical sources \citep{1993zhang,2002Harmon,2004Harmon,2012Wilson-Hodge,2014Rodi,2021Singhal}. 
As near-Earth satellites with specific inclinations and trajectories enter the Earth's occultation zone, the fluxes from certain sources exhibit a stepwise decay due to atmospheric attenuation near the Earth's surface. This decay process reflects the variation in atmospheric density at different altitudes. By fitting the occultation light curve of an X-ray source, it is possible to derive atmospheric density at various altitudes. For X-ray photons, interactions are primarily limited to atomic electrons \citep{2007Determan,2021Katsuda,2023Xue}, making the occultation of stable X-ray sources a reliable method for retrieving neutral atmospheric density in the middle and upper layers.

Neutral atmospheric density plays a crucial role in space engineering as an environmental parameter of the middle and upper atmosphere, such as in determining satellite orbits \citep{2005Storz} and predicting the landing point of reentry \citep{2021Fedele}. However, measuring atmospheric density within the altitude range of 50\,–-300\,km remains challenging, as data in this region are sparse, leading to the so-called 'Thermosphere gap". 
The use of X-ray occultation data offers a promising and cost-effective solution to bridge this gap, enabling precise atmospheric density measurements without requiring dedicated observational time.
For instance, \cite{katsuda2024} utilized observations of the Cassiopeia A supernova remnant obtained from the \textit{Insight}-HXMT telescope in 2022 to measure atmospheric density within the altitude range of 90\,--150\,km through Earth occultation before and after a volcanic eruption.
This approach demonstrates the potential of EOT for advancing our understanding of atmospheric density in this critical altitude range.

Currently, neutral atmospheric density distributions are predominantly obtained from empirical atmospheric models. 
Among these, the Navy Research Laboratory Mass Spectrometer Incoherent Scatter Radar Exosphere (NRLMSISE) model, developed by the U.S. Naval Research Laboratory, is one of the most widely used empirical models for the Earth's atmosphere\citep{1987MSIS,2002MSIS,2021MSIS}. It provides a parameterized description of the average spatiotemporal variation of atmospheric state variables, integrating physical constraints with empirical data. The model inputs include geographic location, time of day, year, and solar and geomagnetic activity, enabling it to simulate atmospheric changes under varying conditions.  The NRLMSIS 2.0 and NRLMSIS 2.1 are updated versions that incorporate enhancements in representing the neutral atmosphere's composition, temperature, and density profiles with altitude. These models are extensively utilized in the scientific community to simulate atmospheric conditions under varying solar and geomagnetic activity levels.
The NRLMSIS series of empirical models \citep{1987MSIS,2002MSIS,2021MSIS} has undergone multiple updates and has become a standard tool in space research. However, its accuracy still requires further validation. Notably, it has been widely demonstrated that the NRLMSISE-00 model tends to overestimate atmospheric densities in the middle and upper atmosphere \citep{2007Determan,2021Katsuda,2022Yu1,2022Yu2,2023Xue}. 
Therefore, validating the newly updated NRLMSIS 2.0 and 2.1 models using X-ray Earth occultation data provides a valuable and independent method to assess their accuracy and reliability.

Using a Bayesian atmospheric density retrieval method combined with EOT to fit the Crab Nebula's occultation light curve, \citet{2023Xue} found that the updated NRLMSIS 2.0 model performs accurately during periods of low solar activity. However, solar activity significantly impacts the Earth's atmosphere \citep{1991Jakowski,kutiev2013,Emmert2015}. \citet{Solomon2014} revealed an unexpectedly weak ionosphere during the prolonged solar minimum between cycles 23 and 24, indicating the reduction of solar extreme ultraviolet (EUV) plays the largest role in causing the ionospheric change. Additionally, \citet{2015Meier} noted that atmospheric density is strongly influenced by the solar cycle, leading to periodic variations that complicate measurements. 
Consequently, the accuracy of the NRLMSIS series models must be rigorously tested during periods of heightened solar activity. In this study, we adopt the method of \citet{2023Xue} to measure atmospheric density and evaluate the performance of the NRLMSIS models during such high-activity periods.

This paper is structured as follows. Section \ref{Method} describes the data collection and reduction process for the Crab Nebula observations by \textit{Insight}-HXMT and presents the maximum likelihood estimation method for retrieving atmospheric density. The results are discussed in Section \ref{result and discussion}, and a summary is provided in Section \ref{summary}.

\begin{figure*}[htb]
    \centering
    \setlength\tabcolsep{0pt}
    \begin{tabular}{cc}
        \includegraphics[width=0.419\textwidth]{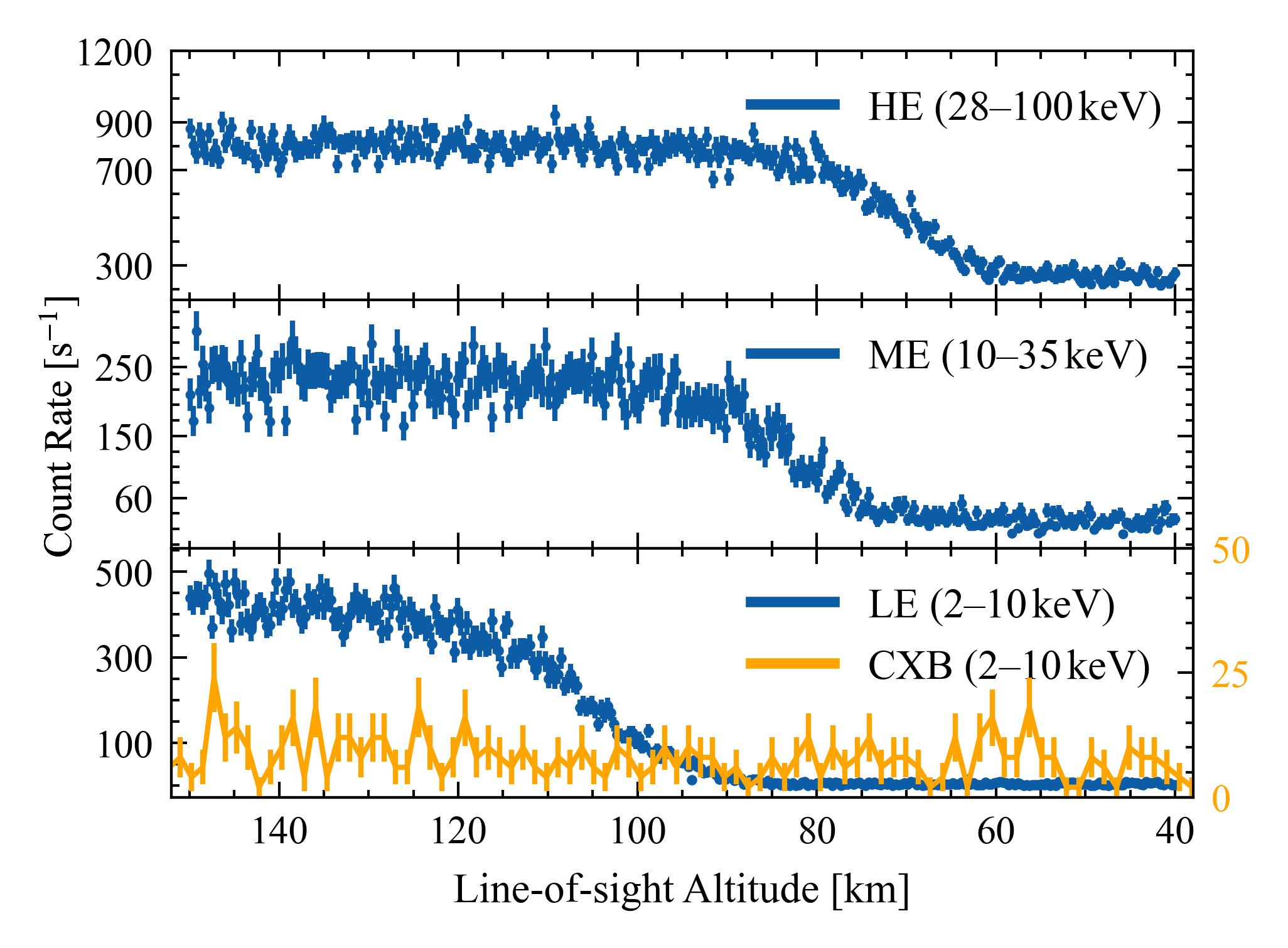}
         & 
        \includegraphics[width=0.439\textwidth]{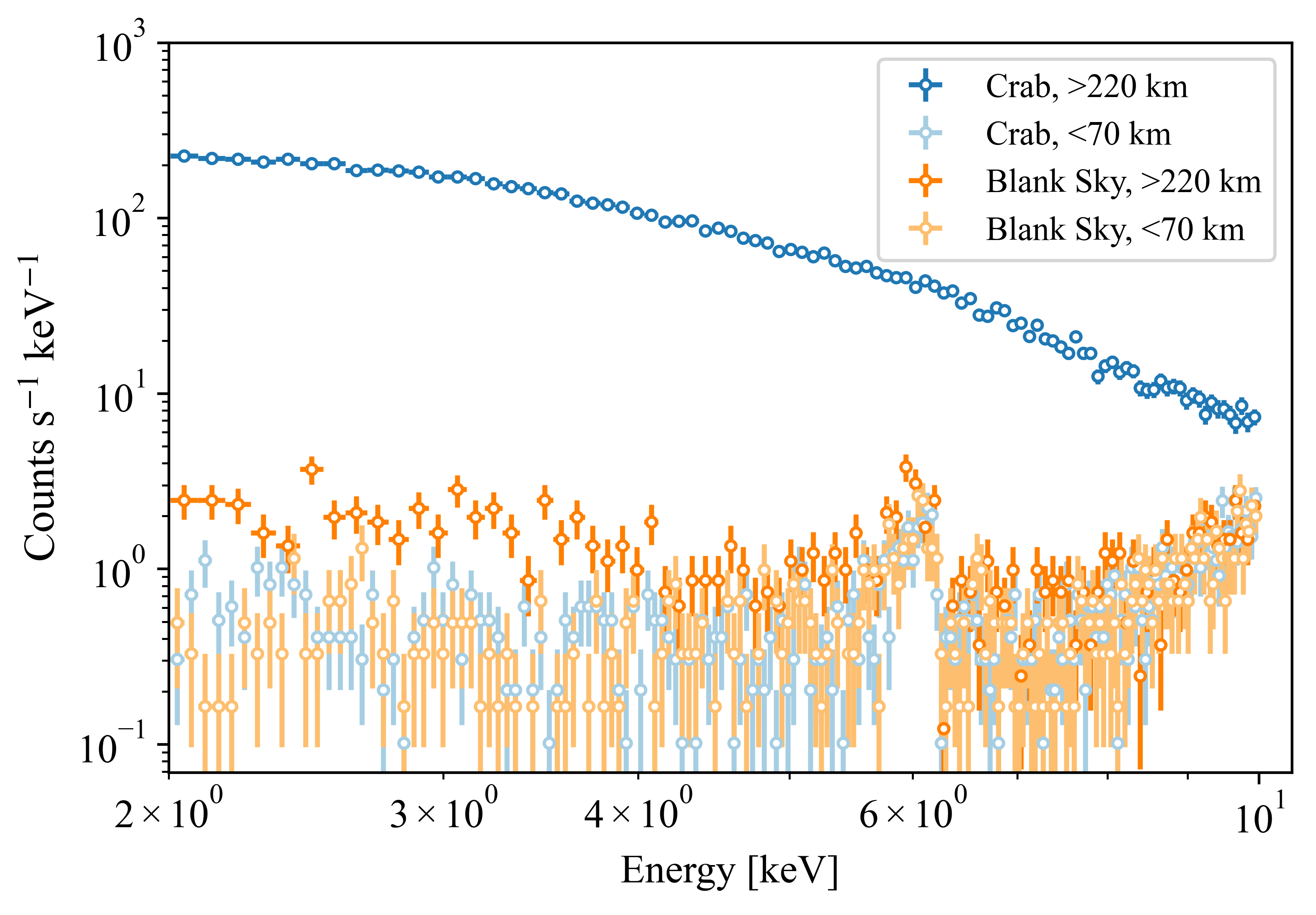}
    \end{tabular}
    
    \caption{\textit{Left}: The X-ray photon counting rate along the LOS  varies with altitude for three telescopes. For LE, ME, and HE telescopes, the observed X-ray photon counting rate is significantly attenuated when the LOS altitude is 90\textendash130\,km, 70\textendash90\,km, and 55\textendash80\,km, respectively. The blue data points and orange solid lines represent the change of Crab and CXB count rates with LOS altitude, respectively.  \textit{Right}: The Spectral Energy Distribution of  Crab and  Blank Sky from 2-10\,keV. Two different shades of color represent complete occultation ($<$70\,km) and without occultation ($>$220\,km). }
    \label{case of occultation}
\end{figure*}

\section{Observations and Methods}
\label{Method}
\subsection{Observations and Data Reduction}
\textit{Insight}-HXMT, China’s first X-ray astronomy satellite, operates at an altitude of 550\,km with an inclination of 43\,degrees \citep{Zhang2020_HXMT}. It is equipped with three main scientific payloads: the High Energy X-ray telescope (HE, 20--250\,keV, 5000 $\rm cm^{2}$, timing resolution: 2\, $\rm{\mu s}$) \citep{Liu2020HE}, the Medium Energy X-ray telescope (ME, 8--35\,keV, 952 $\rm cm^{2}$, timing resolution: 6\, $\rm{\mu s}$) \citep{Cao2020ME} and the Low Energy X-ray telescope (LE, 1--10\,keV, 384 $\rm cm^{2}$, timing resolution: 1\, $\rm{\mu s}$) \citep{Chen2020LE}. 
With its broad energy band, large effective areas, and high time resolution \citep{Li2020InflightCalibration}, \textit{Insight}-HXMT is particularly well-suited for measuring neutral atmospheric density in the middle and upper layers of the Earth using Earth occultation data \citep{2023Xue}. The Crab Nebula, one of the standard calibration sources for \textit{Insight}-HXMT, has been extensively observed, resulting in a substantial accumulation of pointed observational data.
To extract the occultation data of the Crab Nebula, we employed the HXMT Data Analysis Software (\texttt{HXMTDAS}). The primary procedures for extracting occultation data align with those described in subsection 2.1 of \cite{2023Xue}. The 2--10\,keV LE data, 10--35\,keV ME data, and 28--100\,keV HE data are used in the following analysis. We excluded the data obtained when the LE telescope is saturated by the bright Earth and when the HE telescope switches to GRB mode. All the occultation data analyzed in this paper are summarized in Appendix \ref{Table}, which also describes how these data are categorized. From the first to the fifth columns represent the ObsID, the UTC when the LOS altitude is 90\,km, the geographical position of the corresponding tangent point, the latitudinal span of the LOS in the altitude range of 40\textendash150\,km, 
the available telescopes during the occultation and the corresponding occultation type. 

\subsection{The Estimation of Background}\label{The Estimation of Background}
Since the background estimate of \texttt{HXMTDAS} is not accurate in the Occultation Time Interval (OTI), OTI is often excluded from the standard Good Time Interval (GTI). As shown by \cite{2023Xue}, more accurate background estimates can be obtained by fitting a time-dependent model. Following the analysis procedure of \cite{2023Xue}, we binned the occultation data into 0.5\,s time scale, generating $N$ counting data $\{D_{c,i}\}$ for energy channel $c$, where $i = 1, \ldots, N$. For each energy channel $c$ , the background count rate $b_{c}(t)$ is supposed to be
\begin{equation}
    b_c(t) = \exp{\left(\alpha_c + \beta_c t\right)}, \label{background evaluate}
\end{equation}
where $\alpha_{c}, \, \beta_{c}$ is the parameter of background count rare for each channel. Then for each channel, the total count rate $m_{c}(t)$ is
\begin{equation}
    m_{c}(t)=b_{c}(t)+\delta(t)\int_{E}R(E,c)I_{\rm{{0}}}(E)dE,\label{spcetral fit}
\end{equation}
where $\delta(t)$ is 1 if source is occulted at time $t$ and 0 otherwise, $I_{\rm 0}(E)$ is the source flux which can be derived from the observation of the most recent standard GTI, and $R(E,c)$ is the response matrix of telescopes. By fitting Equation (\ref{spcetral fit}) to the light curve outside the OTI, we obtained the estimated value of $b_c(t)$, which will be interpolated into OTI in later analysis. The data whose background cannot be well fitted by Equation (\ref{background evaluate}) is excluded from further analysis. The procedure to assess the background model fitness is detailed in \cite{2023Xue}.

 The left panel of Figure \ref{case of occultation} shows that the count rate of the light curves from three different telescopes decreases with decreasing line-of-sight altitude. The counts spectrum of Crab and the blank sky in right panel of Figure \ref{case of occultation} show that the Cosmic X-ray Background (CXB) is much smaller than the flux of Crab. Therefore, the 2--10\,keV background variation originating from CXB occultation is quite smooth when compared to Crab occultation, and can be described by Equation (\ref{background evaluate}). The LE data we analyzed is down to 2\,keV, which is lower than that of 6\,keV in the previous study \citep{2023Xue}.

\begin{figure*}[htb]
    \centering
    \setlength\tabcolsep{0pt}
    \begin{tabular}{cc}
        \includegraphics[width=0.393\textwidth]{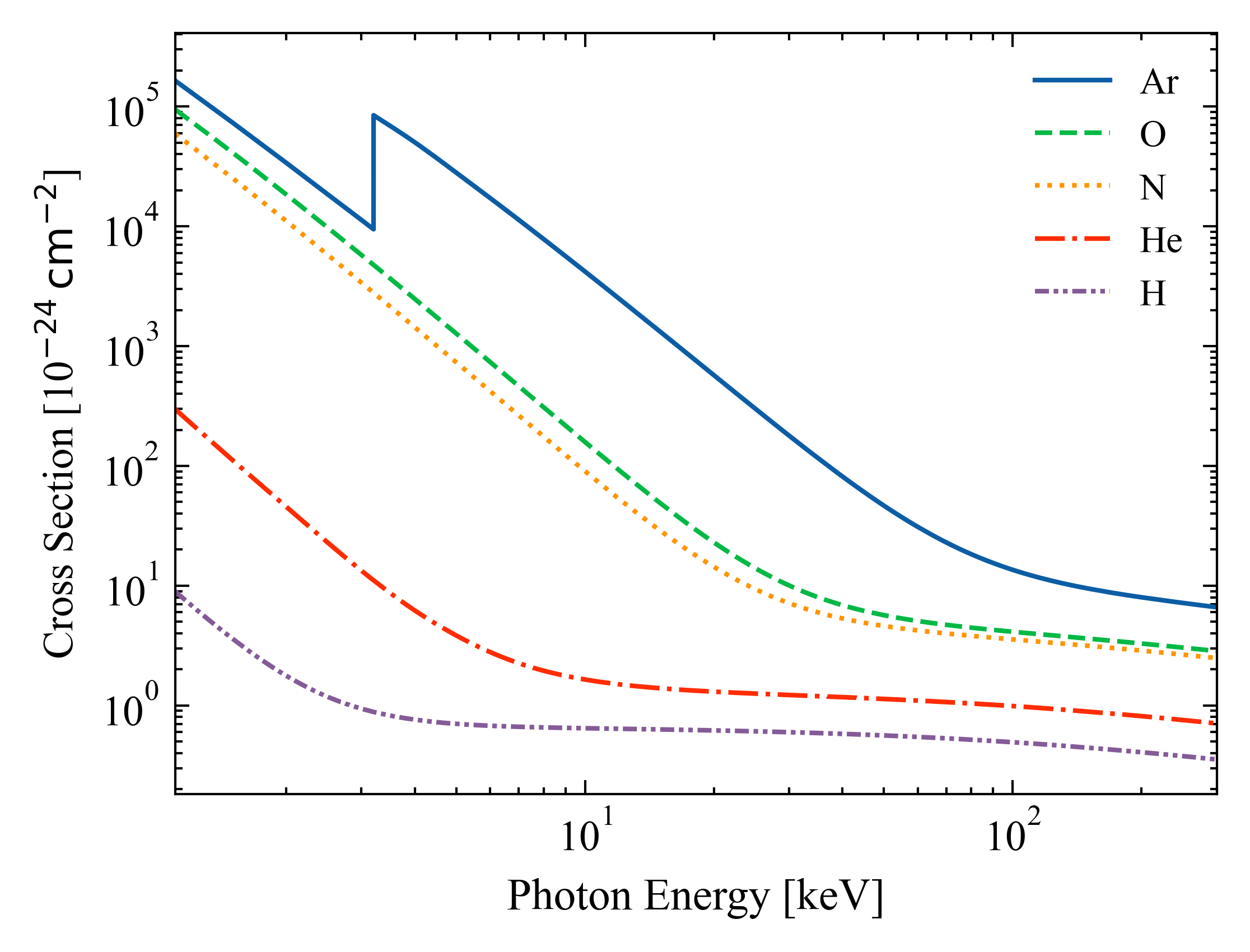}
         & 
        \includegraphics[width=0.4\textwidth]{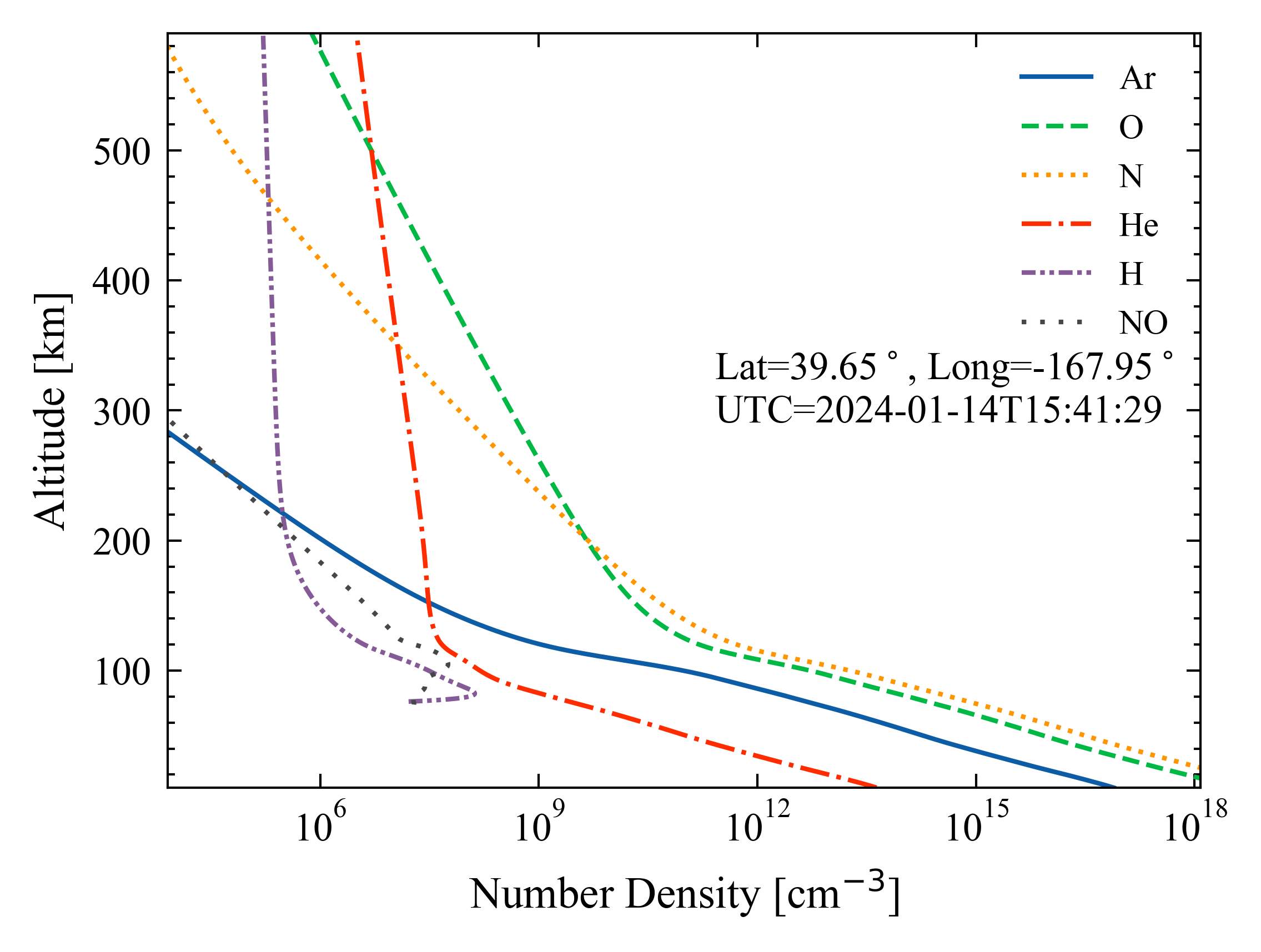}
    \end{tabular}
    \caption{ \textit{Left}: The total X-ray cross-section of five kinds of atoms, including the cross sections of Compton scattering and photo-electric absorption. \textit{Right}: The vertical density profile of five kinds of atoms and NO obtained from NRLMSIS 2.1. Here, the density profile of O includes the contribution from O, O$_2$, and Anomalous O, while the density profile of N includes the contribution from N and N$_2$. }\label{vertical density and cs}
\end{figure*}
\subsection{The Atmospheric Density Retrieval} 

When occultation occurs, X-ray photons of different energies of the source will travel through the atmosphere at different altitudes, which can be used to retrieve the atmospheric density at different altitudes. The three telescopes onboard HXMT, i.e., LE, ME, and HE are sensitive to X-ray photons in different energy ranges, and the occultation data of these telescopes can be used to retrieve the atmospheric density at different altitudes. Figure \ref{case of occultation} shows how the X-ray photon count rate of the Crab varies with altitude during the occultation
for three X-ray telescopes of \textit{Insight}-HXMT. It can be found that HE, ME, and LE count rates decreased significantly within the LOS altitudes of 55-80 km, 70-90 km, and 90-130 km respectively, which means that the occultation data can be used to retrieve the atmospheric densities at corresponding altitudes. To maximize the use of the data for each occultation and improve the photon statistics, we stratify the atmosphere according to the telescopes available for each occultation. By jointly analyzing the occultation data of three telescopes, we can retrieve atmospheric density at multiple altitudes simultaneously. The criteria of atmospheric stratification are shown in Table \ref{stratification}. For example, if only LE telescopes are available for an occultation, the atmospheric densities above 90 km altitude will be retrieved for this occultation.

The attenuation of X-ray photons in the atmosphere can be approximated by the Beer-Lambert law 
\begin{equation}
    I(E,t) = I_0(E)\,\exp{\left[-\tau(t,E)\right]},\label{beer-lambert law}
\end{equation}
where $I(E,t)$ is the  attenuation flux, $I_{\rm 0}(E)$ is the source flux, and $\tau(t,E)$ is the optical depth for LOS, which can be given by
\begin{equation}
    \tau(E,t) = \sum_{s} \sigma_{s}(E)\,N_{s}(t), \label{tau}
\end{equation}
where $\sum_{s} \sigma_{ s}(E)\,N_{s}(t)$ is the sum of the X-ray cross section $\sigma_{s}$ and column density  $N_{s}(t)$ of each atom species $s$. $N_{s}$ is defined as 
\begin{equation}
        N_{s}(t) = \int_{l_{0}(t)}^{l_{*}}n_{s}(l)\,\mathrm{d}l,\label{column density}
\end{equation}
where $l_{0}(t)$ is the satellite position at time $t$, $l_{*}$ is the source position, and $n_{s}(l)$ is the number density of atom species $s$ at location $l$ along the LOS. From the above equation, we know that atmosphere attenuation depends on $n_{s}(l)$, which can be obtained in NRLMSIS series \citep{1987MSIS,2002MSIS,2021MSIS,2022JMSIS}. Our fitting for $n_{s}(l)$ will start from a basic density given by the NRLMSIS model. In other words, we multiply an overall density correction factor $\gamma$ for the number density obtained from the NRLMSIS model. Then, we have
\begin{equation}
    n_{s}(lat_l, lon_l, h_l) = \gamma(h_l)\,n^{\prime}_{s}(lat_l, lon_l, h_l),
\end{equation}  

where $lat_l$, $lon_l$ and $h_l$ are the latitude, longitude and altitude at the location $l$, $n^{\prime}_{s}$ is the density given by the NRLMSIS model, $
\gamma$  is the overall density correction factor for all atoms specie $s$ and is a piecewise function of altitude. $N_{\gamma}$ is the component number of $\gamma$, which is dependent on the altitude boundaries listed in Table \ref{stratification}. For example, if LE, ME, and HE telescopes are all available for
an occultation, the correction factor ${\gamma}$ will have $N_\gamma=14$ components,
\begin{equation}
    \gamma(h) = 
    \begin{cases}
        \gamma_1, & \text{if 55\,km $\leq h < $ 65\,km},\\
        \gamma_2, & \text{if 65\,km $\leq h <$ 70\,km},\\
        \gamma_3, & \text{if 70\,km $\leq h <$ 75\,km},\\
        \gamma_4, & \text{if 75\,km $\leq h <$ 80\,km},\\
        \gamma_5, & \text{if 80\,km $\leq h <$ 85\,km},\\
        \gamma_6, & \text{if 85\,km $\leq h <$ 90\,km},\\
        \gamma_7, & \text{if 90\,km $\leq h <$ 95\,km},\\       \gamma_8, & \text{if 95\,km $\leq h <$ 100\,km},\\
        \gamma_9, & \text{if 100\,km $\leq h <$ 105\,km},\\
        \gamma_{10}, & \text{if 105\,km $\leq h <$ 110\,km},\\
        \gamma_{11}, & \text{if 110\,km $\leq h <$ 115\,km},\\
        \gamma_{12}, & \text{if 115\,km $\leq h <$ 120\,km},\\
        \gamma_{13}, & \text{if 120\,km $\leq h \leq$ 130\,km},\\
        \gamma_{14}, & \text{if 130\,km $\leq h \leq$ 550\,km}.
    \end{cases}
\end{equation}
which makes left side of Equation (\ref{beer-lambert law})  replace to $I(\gamma,E,t)$, and Equation (\ref{tau}), (\ref{column density}) also add the correct factor $\gamma$ accordingly. \\
We also show the  X-ray cross section of H, He, N, O, Ar, and the corresponding vertical density profiles in the left panel of  Figure \ref{vertical density and cs}. The X-ray cross-section of atoms is obtained from \texttt{XCOM} database \citep{1987Berger}, the input environmental parameters for the NRLMSIS model (i.e. 10.7\,cm solar radio flux $F10.7$ and geomagnetic indices $Ap$) are obtained from online database \citep{2021Matzka}. In the left panel of Figure \ref{vertical density and cs}, it can be found that the cross sections of N and O are basically the same over the energy range we are considered, which means that they cannot be distinguished from the light curve of occultation but get their sums. Furthermore,  following the previous work \citep[e.g.,][]{Kaastra2017,2023Xue}, we only consider the atom species of N, O, and Ar. Other atom species only have a little attenuation for X-ray photons because their scattering cross-sections and number densities are both significantly lower than those of N, O, and Ar within 55-130 km altitude (see the right panel of Figure \ref{vertical density and cs}). It is worth mentioning that nitric oxide (NO) density is introduced in the latest NRLMSIS 2.1 model but contributes little to the overall optical depth in the specified altitude range due to its low abundance. The counts from the source of each channel $c$ can be modeled as
\begin{equation}
    S_{c,i}(\gamma) = \frac{\Delta t_i}{\Delta Ti}\int_{\Delta {Ti}} \mathrm{d}t \int_E \mathrm{d}E \ R(E,c) \, I(\gamma,E,t).
\end{equation}

During $\Delta T_i$, we know the observed counts $D_{c,i}$, and the background estimate $\hat{B}_{c,i}$ and corresponding error $\sigma_{c,i}$ obtained in Section \ref{The Estimation of Background}. The likelihood for Poisson count data with Gaussian background is
\begin{equation}
\begin{split}
    L\left(S_{c,i}(\gamma) \, \middle| \, D_{c,i},\hat{B}_{c,i},\sigma_{c,i} \right) =  L\left(D_{c,i}\, \middle| \, S_{c,i}(\gamma),B_{c,i} \right) \\ 
    \times \, L\left(B_{c,i} \, \middle| \, \hat{B}_{c,i},\sigma_{c,i} \right) \, \label{likelihood},
\end{split}
\end{equation}

where
\begin{equation}
\begin{split}
    L\Bigl(D_{c,i}\, \big| \, S_{c,i}(\boldsymbol{\gamma}),B_{c,i} \Bigr) = \frac{\left[S_{c,i}(\boldsymbol{\gamma})+B_{c,i}\right]^{D_{c,i}}}{D_{c,i}!}
    \\
    \times\exp{\bigl[-\left(S_{c,i}(\boldsymbol{\gamma})+B_{c,i}\right)\bigr]},
\end{split}
\end{equation}
\begin{equation}
        L\left(B_{c,i} \, \middle| \, \hat{B}_{c,i},\sigma_{c,i} \right) = \frac{1}{\sqrt{2\pi}\sigma_{c,i}}\exp{\left[-\frac{\left(B_{c,i}-\hat{B}_{c,i}\right)^2}{2\sigma_{c,i}^2}\right]}\,,
\end{equation}
and $B_{c,i}$ is the profiled background \citep[see, ][]{2022XSPEC, 2023Xue}.

Finally, the likelihood of observing the data $\boldsymbol{D}$ during the OTI can be given as
\begin{equation}
\begin{split}
    L\left(\boldsymbol{S}(\boldsymbol{\gamma}) \, \middle| \, \boldsymbol{D},\hat{\boldsymbol{B}},\boldsymbol{\sigma} \right) =\prod\nolimits_{1 \leq i \leq N,\,\Delta T_i \in \mathrm{OTI}} \prod\nolimits_{c_{\mathrm{min}} \leq c \leq c_{\mathrm{max}}} \\ L\left(S_{c,i}(\boldsymbol{\gamma}) \, \middle| \,D_{c,i} ,\hat{B}_{c,i},\sigma_{c,i} \right) \,.
\end{split}
\end{equation}

The best fit $\boldsymbol{\gamma}$ can be obtained by maximizing the above likelihood.

\begin{figure*}[htb]
    \centering
    \setlength\tabcolsep{0pt}
    \begin{tabular}{cc}
        \includegraphics[width=0.5\textwidth]{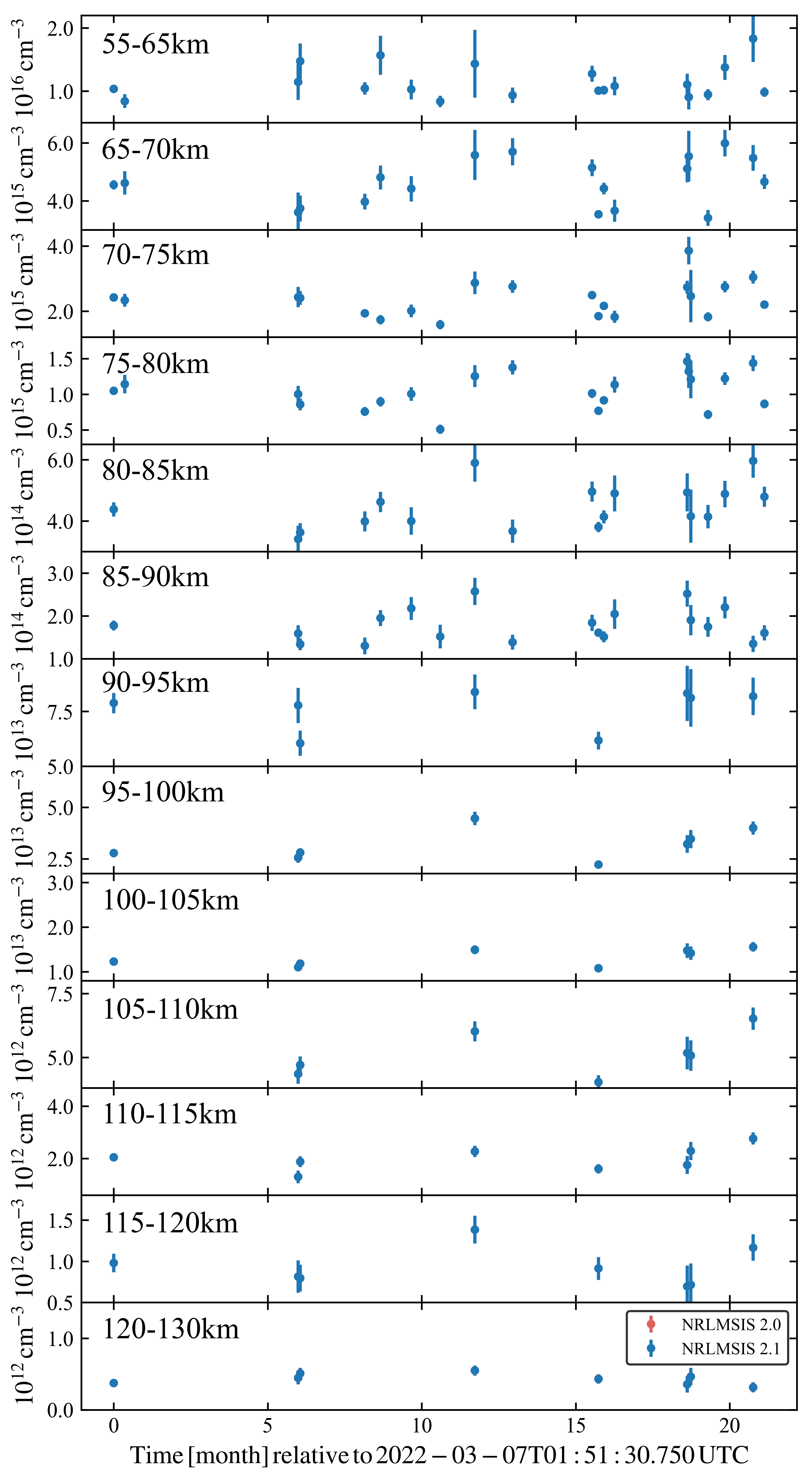}
         & 
        \includegraphics[width=0.5\textwidth]{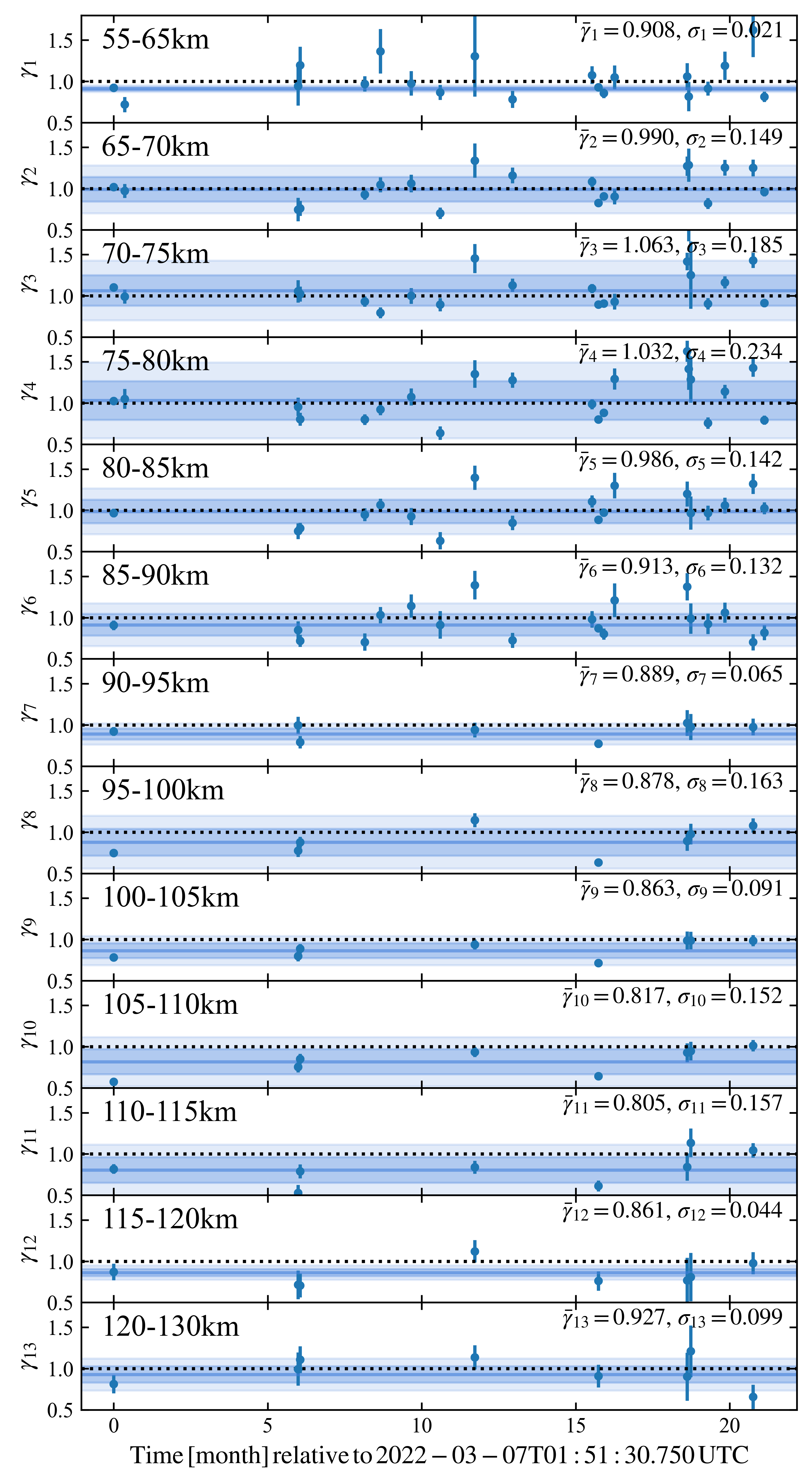}
    \end{tabular}
    \caption{The inversion result of N+O atmospheric densities (left panel) and the density correction factors (right panel) at altitudes of 55–130 km, based on the NRLMSIS 2.0 (red points) and NRLMSIS 2.1 (blue points) models. The solid lines with different transparencies represent the central value, the 1-$\sigma$ error line, and the 1.96-$\sigma$ error line (95\% confidence interval), respectively. The $\bar{\gamma}$ and $\sigma$ are determined by MLE (see text). 
 For clarity, the retrieved densities of observation P0502132014 (the eleventh data point) are shifted left by 4 months.} \label{result}
\end{figure*}
\begin{figure*}[htb]
    \centering
    \setlength\tabcolsep{0pt}
    \begin{tabular}{cc}
        \includegraphics[width=0.41\textwidth]{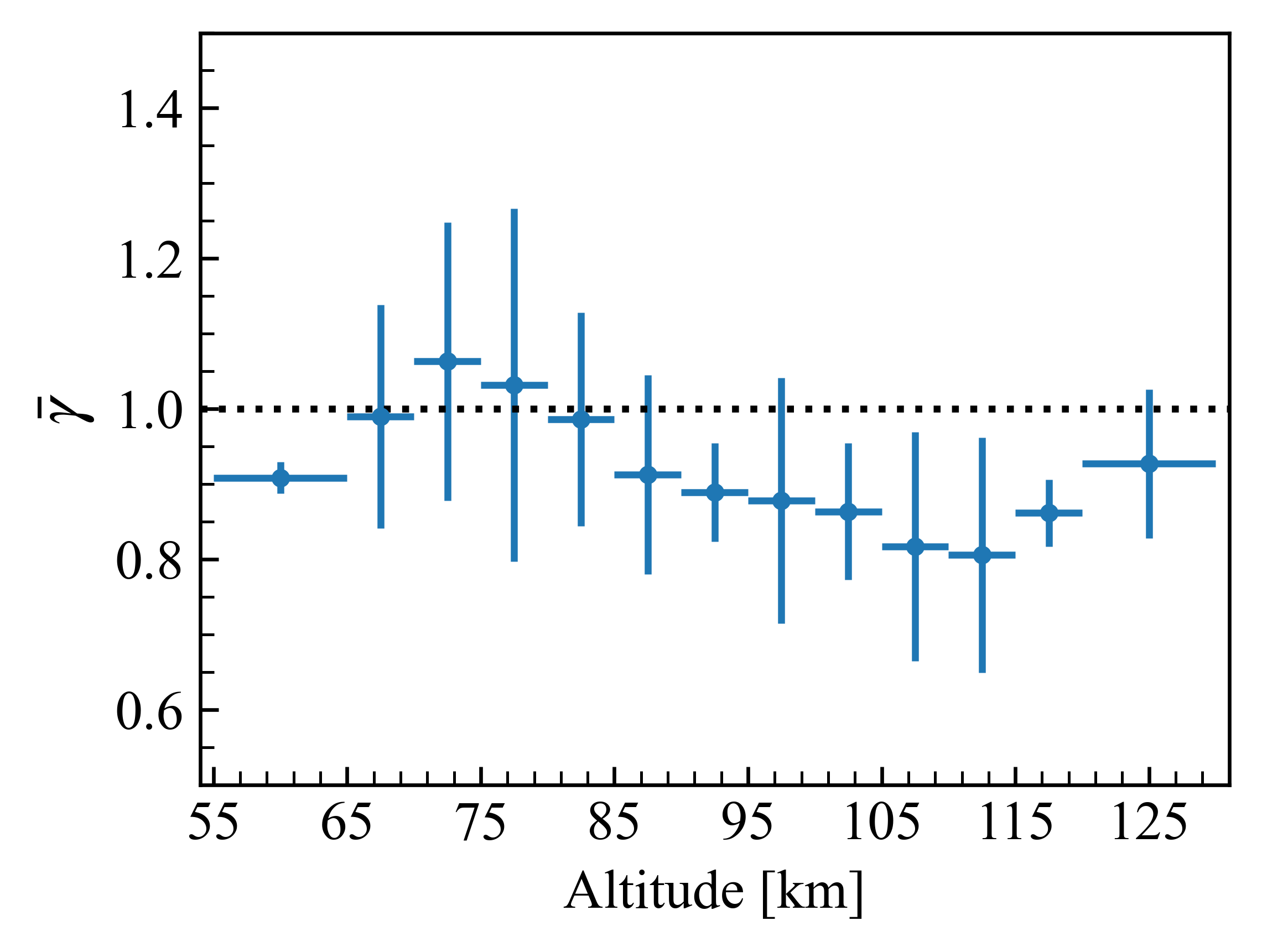}
         & 
        \includegraphics[width=0.4\textwidth]{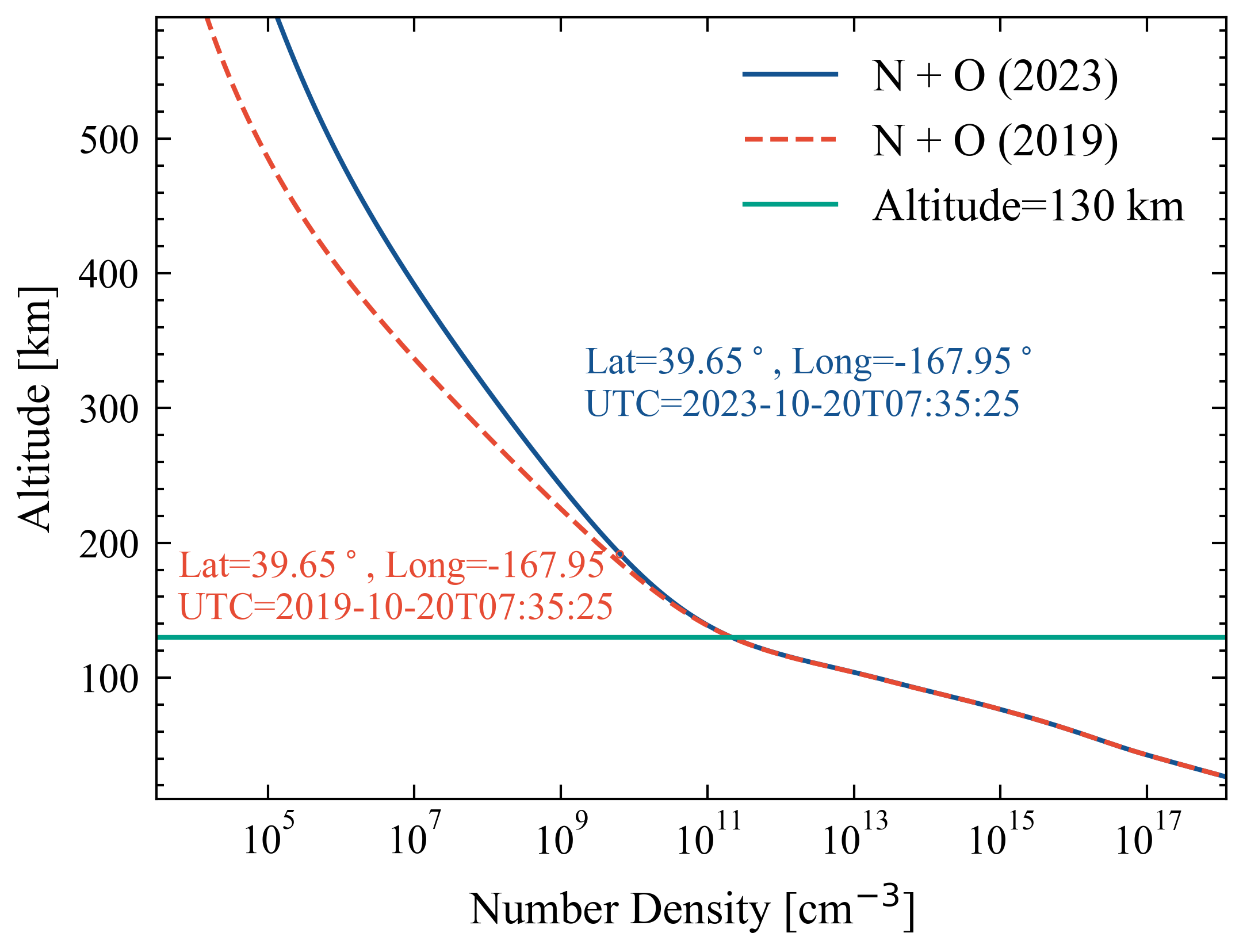}
    \end{tabular}
    \caption{ \textit{Left}: The blue points represent the mean correction factors $\bar{\gamma}$  at altitudes of 55–130 km with 1--$\sigma$ error bar, using the NRLMSIS 2.1 model.  \textit{Right}: Comparison of N+O atmospheric density predicted by the NRLMSIS 2.1 model at the same latitude and longitude but in different years. The blue solid line and the red dotted line represent the distributions for 2023 and 2019, respectively. The green solid line indicates the upper limit altitudes of the inversion in this work. }\label{vertical density and compare}
\end{figure*}

\section{Results and Discussion }
\label{result and discussion}
The analysis of atmospheric density profiles derived from X-ray occultation observations of Insight-HXMT is presented, with a focus on the expanded energy range of 2--10\,keV for LE. This extension enables the retrieval of atmospheric density up to an altitude of 130\,km, as the contribution of photons from CXB to the occultation of Crab Nebula is negligible in this regime as shown in Figure \ref{case of occultation}.
Using the density retrieval methodology outlined in Section \ref{Method}, we analyzed a total of 83 occultation from 21 observation, as detailed in Table \ref{TabA1}. The retrieved atmospheric densities and corresponding correction factors are presented in Figure \ref{result}. To quantify the systematic deviations between the observed densities and model predictions, we employed MLE to determine the mean correction factor $\bar{\gamma}$ and its intrinsic scatter $\sigma$ for each atmospheric layer. The likelihood function is given by:
\begin{equation}
    \prod_i L(\bar{\gamma}|\gamma_i, \sigma_i,\sigma) = \prod_i \frac{1}{\sqrt{2\pi (\sigma_i^2 + \sigma^2)}}\exp\left[-\frac{(\gamma_i-\bar{\gamma})^2}{2(\sigma_i^2+\sigma^2)}\right],
\end{equation}
where $\gamma_i$ is the factor and $\sigma_i$ is the corresponding measurement error of the individual density retrieval.

Our results reveal significant discrepancies between the retrieved atmospheric densities and the predictions of the NRLMSIS 2.0 and 2.1 models. Specifically, in the altitude range of 
90 -- 95 \,km, 100 -- 120 \,km  both models overestimate the atmospheric density by approximately 20\%. This overestimation is consistent with the findings of \cite{katsuda2021}, who reported a similar bias in the N+O density within this range. However, our analysis diverges from theirs in the 70 -- 90\,km and 95 -- 100 \,km altitude range, where the retrieved densities are in good agreement with the NRLMSIS models. This suggests that the systematic overestimation by the NRLMSIS models is altitude-dependent and not uniformly distributed across the upper atmosphere. 
The left panel of Figure \ref{vertical density and compare} illustrates the verification results of the NRLMSIS model at an altitude range of 55 to 130\,km, highlighting the overestimation by the NRLMSIS models at 90 -- 95\,km and 100--120\,km. 

It is worth noting that the atmospheric density at an altitude of 55-65 km seems to have been highly overestimated by NRLMSIS model. We suggest this is likely due to selection bias in the limited samples and can be addressed by incorporating more observations. For example, if another measurement in this altitude range, such as $1.1\pm0.1$, were included, then the mean correction factor would be $0.92\pm0.05$. These findings underscore the need for refinements in atmospheric models, particularly in regions where discrepancies with observational data are pronounced. Such improvements are critical for advancing our understanding of atmospheric composition and dynamics, as well as for enhancing the accuracy of occultation-based density retrievals.

From the left panel of Figure \ref{result}, it is evident that the atmospheric density within the altitude range of 55 -– 130\,km exhibits minimal sensitivity to the increasing solar activity over recent years. This suggests that, at these altitudes, the atmospheric density is relatively stable and not significantly influenced by variations in solar flux. However, the right panel of Figure \ref{vertical density and compare} reveals a contrasting behavior at higher altitudes. Specifically, the comparison of atmospheric density profiles at the same latitude and longitude for 2023 and 2019 indicates that the density above 200 km may be substantially affected by increasing solar activity. This is consistent with the known dependence of thermospheric density on solar irradiance, particularly in the upper atmosphere.

In principle, X-ray photons with energies as low as 1\,keV can be used to retrieve atmospheric density at altitudes of approximately 200\,km, as demonstrated by \cite{yu2022new}. However, the current study is limited by the instrumental constraints of the \textit{Insight}-HXMT mission, which has a minimum detectable energy threshold of 2\,keV. This limitation restricts the sensitivity of our analysis to density variations at altitudes beyond 200\,km, where the influence of solar activity becomes more pronounced.
To address this limitation in future work, we propose utilizing occultation data from missions with lower energy coverage, such as the Einstein Probe\,(EP) (0.5 -- 10\,keV). It is capable of probing atmospheric density variations at altitudes exceeding 200 km, owing to their enhanced sensitivity to lower energy X-ray photons. By leveraging these datasets, we aim to rigorously test the accuracy of the NRLMSIS model in predicting atmospheric density at higher altitudes. Additionally, this approach will enable us to investigate whether the NRLMSIS model adequately captures the expected increase in atmospheric density above 200\,km due to elevated solar activity.

Such advancements will not only refine our understanding of the thermospheric response to solar forcing but also improve the predictive capabilities of atmospheric models, particularly in the context of space weather and its impact on satellite operations and orbital dynamics.

\section{Summary}
\label{summary}
EOT has been demonstrated to be a highly effective and cost-efficient method for retrieving the density profiles of the middle and upper atmosphere. In this study, we employ the Maximum Likelihood Estimation method to analyze occultation data from the Crab Nebula, as observed by the \textit{Insight}-HXMT mission, to derive neutral atmospheric density profiles. By examining the light curve variations and energy spectra of the CXB, we establish that the contribution of CXB photons to the atmospheric density inversion is negligible. This finding allows us to extend the energy range of LE to 2–-10\,keV, enabling the successful retrieval of atmospheric density at altitudes ranging from 55 to 130\,km.

Our analysis of atmospheric density data from 2022 to 2024, obtained through occultation observations, reveals that the NRLMSIS 2.0 and NRLMSIS 2.1 models provide accurate density estimates at altitudes of 65 –- 90\,km, 95 -– 100\,km and 120 -– 130\,km. However, these models overestimate the atmospheric density in the 90 -– 95\,km and 100 -– 120\,km altitude range by approximately 20\%. The atmospheric measurement results within the altitude range of 55 -- 65\,km may contain the selection effect. Therefore, we do not rule out the accuracy of the NRLMSIS prediction within this atmospheric range. Additionally, our results indicate that solar activity has minimal influence on atmospheric density within the 55 -– 130\,km altitude range, as no significant variations were observed over the study period.

For altitudes above 130\,km, particularly in the region beyond 200\,km where solar activity is expected to have a more pronounced impact, further validation of the NRLMSIS model predictions will require data from other soft X-ray observatories, such as the Einstein Probe (EP). They will be probe these higher altitudes and assess the accuracy of atmospheric density predictions under varying solar activity conditions. Such efforts will enhance our understanding of thermospheric dynamics and improve the reliability of atmospheric models for applications in space weather forecasting and satellite operations.

\section*{acknowledgements}

This work made use of data from the \textit{Insight}-HXMT mission, funded by the China National Space Administration and CAS. 
The authors thank supports from National Key R\&D Program of China (Grant No. 2021YFA0718500), the National Natural Science Foundation of China (Grant Nos. 12273043, 12393811, 
12273042, 12494572), 

and the Strategic Priority Research Program of the Chinese Academy of Sciences (Grant Nos. E02212A02S, XDB0550300).

\bibliography{sample631}{}
\bibliographystyle{aasjournal}


\appendix

\section{Occultation Data List}
\centering
\begin{center}
\renewcommand{\arraystretch}{0.995}
\begin{longtable}{cccccc}

\caption{Summary of occultation data analyzed in this paper.}\label{TabA1}\\
\hline
ObsID & UTC$^a$ & \begin{tabular}{c}Tangent point$^b$ \\ Lat, Long [$^\circ$] \end{tabular} & Lat span$^c$ [$^\circ$] & Telescope & Type$^d$ \\
\hline
\endfirsthead
\hline
ObsID & UTC$^a$ & \begin{tabular}{c}Tangent point$^b$ \\ Lat,\,Long [$^\circ$] \end{tabular} & Lat span$^c$ [$^\circ$] & Telescope & Type$^d$ \\
\hline
\endhead
\hline

\endfoot
\hline
\multicolumn{6}{l}{\small $^a$UTC time when LoS altitude $h = 90$\,km.}\\
\multicolumn{6}{l}{\small $^b$Latitude and longitude of the tangent point at $h = 90$\,km.}\\
\multicolumn{6}{l}{\small $^c$Latitudinal span of LoS, within altitude 40\textendash150\,km.}\\
\multicolumn{6}{l}{\small $^d$The letter S stands for Setting and the letter R stands for Rising.}\\
\endlastfoot
P0402349009 & 2022-03-07T01:52:41 & 17.55,\,-11.60 & (5.60,\,27.00) & ME+HE & S \\
 & 2022-03-07T08:14:46 & 20.38,\,-106.09 & (8.05,\,29.90) & LE+ME & S \\
 & 2022-03-07T09:50:07& 21.11,\,-121.69 & (8.69,\,30.65) & HE & S \\
 & 2022-03-07T11:25:38 & 21.84,\,-153.29 & (9.30,\,31.37) & HE & S \\
 & 2022-03-07T13:01:09 & 22.54,\,-176.90 & (9.91,\,32.13) & ME+HE & S \\
 & 2022-03-07T14:36:40 & 23.27,\,159.50 & (10.52,\,32.85) & LE+ME & S \\
 & 2022-03-07T16:12:11 & 23.97,\,135.90 & (11.15,\,33.61) & LE+ME+HE & S \\

 \hline
 P0402349010 & 2022-03-17T18:23:54 & -16.30,\,-90.53 & (-24.78,\,-5.98) & HE & R \\
 & 2022-03-17T19:59:18 & -16.61,\,-114.31 & (-25.09,\,-6.29) & HE & R \\
\hline
P0502132001 & 2022-09-02T16:50:37 & 38.05,\,101.04 & (25.69,\,45.43) & ME & S \\
 & 2022-09-02T18:25:53 & 38.23,\,77.03 & (25.82,\,45.62) & LE & R \\
 & 2022-09-02T21:36:23 & 38.58,\,29.02 & (26.11,\,46.00) & ME & R \\
 & 2022-09-02T23:11:39 & 38.76,\,5.02 & (26.27,\,46.18) & LE+ME+HE & R \\
\hline
 P0502132002 & 2022-09-04T16:28:41 & 43.05,\,100.82 & (29.75,\,50.81) & ME+HE & R \\
 &2022-09-04T18:03:57 & 43.22,\,76.80 & (29.87,\,50.98) & LE+ME & R \\
   & 2022-09-04T19:39:10 & 43.37,\,52.80 & (30.00,\,51.14) & ME & R \\
 &2022-09-04T21:14:26 & 43.52,\,28.79 & (30.11,\,51.32) & LE+ME+HE & S \\

 &2022-09-04T22:49:42 & 43.68,\,4.78 & (30.24,\,51.49) & LE+ME & S \\
\hline
 P0502132004 & 2022-11-06T11:36:02 & 36.35,\,116.98 & (24.27,\,43.64) & HE & R \\
 &2022-11-06T13:11:16 & 36.54,\,92.98 & (24.43,\,43.83) & ME+HE & R \\
 
   & 2022-11-06T14:46:30 & 36.72,\,68.98 & (24.57,\,44.03) & ME+HE & S \\
 &2022-11-06T16:21:44 & 44.99,\,36.91 & (24.73,\,44.22) & ME & R \\
  & 2022-11-06T17:56:56 & 37.09,\,20.99 & (24.89,\,44.41) & ME+HE & S \\
 &2022-11-06T19:32:10 & 37.27,\,-3.00 & (25.03,\,44.61) & ME+HE & R \\

\hline
 P0502132005 & 2022-11-21T22:31:38 & -0.13,\,135.13 & (-10.00,\,9.43) & ME+HE & S \\
 &2022-11-22T00:06:59 & 0.43,\,111.45 & (-9.52,\,9.96) & ME & S \\
  &2022-11-22T04:53:05 & 2.05,\,40.39 & (-8.04,\,11.56) & ME & S \\

\hline
 P0502132007 & 2022-12-21T16:57:06 & -160.73,\,22.82 & (12.58,\,29.83) & ME+HE & S \\
 &2022-12-21T18:32:20 & 22.58,\,175.28 & (12.37,\,29.59) & ME+HE & S \\
  &2022-12-21T21:42:45 & 22.11,\,127.31 & (11.96,\,29.13) & ME & S \\
\hline
P0502132008 & 2023-01-19T02:21:32 & 54.08,\,166.42 & (38.09,\,63.57) & ME+HE & S \\
 & 2023-01-19T03:56:42 & 54.18,\,142.41 & (38.17,\,63.68) & ME+HE & R \\

\hline
P0502132009 & 2023-02-21T15:58:51 & 33.70,\,158.60 & (22.03,\,40.87) & ME+HE & S \\
 & 2023-02-21T22:19:30 & 32.91,\,62.71 & (21.36,\,40.06) & ME & S \\
 & 2023-02-22T01:29:50 & 32.50,\,14.75 & (21.02,\,39.64) & ME & S \\

\hline
P0502132011 & 2023-03-30T11:45:38 & -14.25,\,164.05 & (-22.94,\,-4.03) & LE & S \\
 & 2023-03-30T13:20:55 & -13.87,\,140.33 & (-22.60,\,-3.71) & LE+HE & S \\

 & 2023-03-30T18:06:44 & -12.78,\,69.15 & (-21.57,\,-2.65) & ME & S \\
  & 2023-03-30T19:42:01 & -12.39,\,45.43 & (-21.23,\,-2.30) & ME+HE & S \\
\hline
P0502132014 & 2023-10-13T19:42:17 & -4.39,\,-145.46 & (-13.91,\,5.31) & ME+HE & S \\
 & 2023-10-13T21:17:30 & -3.92,\,-169.13 & (-13.46,\,5.78) & ME+HE & S \\
 & 2023-10-14T00:27:49 & -2.92,\,143.53 & (-12.56,\,6.75) & ME & S \\
 & 2023-10-14T02:03:02 & -2.42,\,119.87 & (-12.12,\,7.21) & HE & S \\
\hline
P0602122001 & 2023-10-20T06:00:09 & 40.32,\,-144.55 & (24.86,\,50.87) & ME+HE & R \\
 & 2023-10-20T07:35:25 & 39.65,\,-167.95 & (24.32,\,50.17) & LE+ME+HE & R \\
 & 2023-10-20T09:10:42 & 38.99,\,168.65 & (23.76,\,49.44) & ME+HE & S \\
 & 2023-10-20T10:45:58 & 38.31,\,145.23 & (23.22,\,48.74) & ME & R \\
  & 2023-10-20T12:21:14 & 37.63,\,121.82 & (22.66,\,48.00) & ME & R \\
\hline
P0602122002 & 2023-10-25T08:14:01 & -8.15,\,-159.65 & (-17.35,\,1.70) & ME+HE & R \\
 & 2023-10-25T09:49:12 & -8.61,\,176.68 & (-17.73,\,1.30) & ME+HE & R \\
  & 2023-10-25T11:24:17 & -9.04,\,153.00 & (-18.14,\,0.88) & ME+HE & R \\
   & 2023-10-25T12:59:28 & -9.46,\,129.31 & (-18.53,\,0.48) & ME+HE & R \\
      &2023-10-25T14:34:39
      & -9.89,\,105.63 & (-18.91,\,0.06) & ME & R \\
         & 2023-10-25T16:09:55 & -10.29,\,81.94 & (-19.30,\,-0.34 & ME+HE & R \\
& 2023-10-25T17:45:06 & -10.72,\,58.26 & (-19.68,\,-0.71) & ME+HE & R \\
& 2023-10-26T04:51:18 & -13.41,\,-107.60 & (-22.18,\,-3.26) & ME+HE & R \\
& 2023-10-26T06:26:28 & -13.78,\,-131.30 & (-22.51,\,-3.63) & ME+HE & R \\
\hline
P0602122003 & 2023-11-05T00:26:05 & 45.57,\,147.86 & (31.73,\,53.58) & ME+HE & S \\
 & 2023-11-05T02:01:09 & 45.43,\,123.90 & (31.61,\,53.43) & ME+HE & S \\
\hline

P0602122011 & 2024-01-14T15:41:29 & 30.73,\,-161.06 & (19.50,\,37.82) & ME+HE & S \\
 & 2024-01-14T17:16:28 & 30.52,\,175.01 & (19.32,\,37.61) & LE+HE & R \\
 & 2024-01-15T02:46:29 & 29.26,\,31.41 & (18.23,\,36.32) & ME & S \\
  & 2024-01-15T05:56:30 & 28.83,\,-16.46 & (17.87,\,35.88) & HE & S \\
\hline
P0602122012 

 & 2024-01-16T05:41:38 & 25.54,\,-15.44 & (14.98,\,32.55) & HE & R \\
\hline
P0602122013 & 2024-01-18T00:26:57 & 19.20,\,58.42 & (9.35,\,26.23) & LE+ME & R \\

\hline
P0602122014 & 2024-02-03T22:25:09 & 30.14,\,-129.47 & (19.00,\,37.21) & ME+HE & R \\
& 2024-02-04T00:00:08 & 30.35,\,-153.40 & (19.18,\,37.42) & ME+HE & R \\
& 2024-02-04T01:35:07 & 30.55,\,-177.33 & (19.36,\,37.64) & ME+HE & R \\
& 2024-02-04T11:05:01& 31.79,\,39.10 & (20.42,\,38.90) & ME & R \\
\hline
P0602122015 
 & 2024-02-20T09:36:14 & -8.00,\,-123.14 & (-17.23,\,1.87) & ME+HE & S \\
 & 2024-02-20T15:56:24 & -6.18,\,142.28 & (-15.55,\,3.61) & ME+HE & S \\
\hline
P0602122016 & 2024-03-18T20:09:28 & 33.03,\,70.02 & (21.46,\,40.18) & LE & S \\
 & 2024-03-18T21:44:24 & 32.83,\,46.10 & (21.30,\,39.98) & LE+ME & S \\
 
 & 2024-03-19T04:04:06 & 32.03,\,-49.57 & (20.61,\,39.15) & ME+HE & S \\
 & 2024-03-19T07:13:59 & 31.62,\,-97.40 & (20.27,\,38.73) & LE+HE & S \\
 & 2024-03-19T13:33:44 & 30.80,\,166.93 & (19.56,\,37.89) & LE+HE & S \\
 & 2024-03-19T15:08:40 & 30.59,\,143.01 & (19.37,\,37.68) & LE+HE & S \\
\hline
P0602122017 & 2024-03-29T17:50:55 & -7.81,\,-98.17 & (-15.75,\,0.48) & ME+HE & R \\
 & 2024-03-29T18:49:51 & -7.52,\,60.82 & (-15.47,\,0.75) & HE & S \\
  & 2024-03-29T19:25:52 & -7.56,\,-122.07 & (-15.52,\,0.71) & ME+HE & R \\
   & 2024-03-29T21:00:45 & -7.31,\,-145.97 & (-15.28,\,0.94) & ME+HE & R \\

        & 2024-03-29T22:35:42 & -7.07,\,-169.88 & (-15.04,\,1.17) & ME+HE & R \\
            & 2024-03-29T23:34:38 & -8.27,\,-10.89 & (-16.19,\,0.07) & HE & S \\
                & 2024-03-30T00:10:38 & -6.82,\,166.22 & (-14.80,\,1.40) & ME+HE & R \\

\end{longtable}
\end{center}
\label{Table}

\section{Atmospheric altitude stratification}
\centering

\begin{center}
\begin{table}[htb]
    \centering
    \caption{The six cases of dividing the atmosphere.}
    \resizebox{\textwidth}{!}{
        \hspace{-12.5em}\begin{tabular}{|c|ccccccccccccccc|}
        \hline
            Telescope & \multicolumn{15}{c|}{Altitude boundaries [km]}\\
        \hline
            LE+ME+HE & 55 & 65 & 70 & 75 & 80 & 85 & 90 & 95&100 & 105&110&115&120&130& 550\\
            LE+ME & & & 70 & 75 & 80 & 85 & 90& 95&100 & 105&110&115&120&130&550\\
            ME+HE & 55 & 65 & 70 & 75 & 80 & 85 & 90 & && & & && &550\\
            ME &  & & 70 & 75 & 80 & 85 & 90 & && & &&  & & 550\\
            HE & 55 & 65 & 70 & 75 & 80 & & & & & && & & &550\\
            LE & && & && &   90 &95 &100&105&110&115&120&130 & 550 \\
        \hline
        \end{tabular}
    }
    \label{stratification}
\end{table}
\end{center}

\end{document}